\documentstyle[12pt,epsf]{article}

\begin{document}

\tolerance=5000

\def\be{\begin{equation}}
\def\ee{\end{equation}}
\def\bea{\begin{eqnarray}}
\def\eea{\end{eqnarray}}
\def\nn{\nonumber \\}
\def\e{{\rm e}}

\def\SEH{S_{\rm EH}}
\def\SGH{S_{\rm GH}}
\def\AdS5{{{\rm AdS}_5}}
\def\S4{{{\rm S}_4}}

\  \hfill
\begin{minipage}{3.5cm}
February 2002 \\
AEI-2003-17 \\
\end{minipage}

\vfill

\begin{center}
{\large\bf The one-loop vacuum energy and RG flow induced by double-trace operators 
in AdS/CFT and dS/CFT correspondence}

\vfill

{\sc Shin'ichi NOJIRI}\footnote{nojiri@cc.nda.ac.jp}
and {\sc Sergei D. ODINTSOV}$^{\spadesuit}
$\footnote{
odintsov@ieec.fcr.es }

\vfill

{\sl Department of Applied Physics \\
National Defence Academy,
Hashirimizu Yokosuka 239-8686, JAPAN}

\vfill

{\sl $\spadesuit$
Lab. for Fundamental Studies,
Tomsk State Pedagogical University,
634041 Tomsk, RUSSIA and ICREA/IEEC, Barcelona,
SPAIN}

\vfill

{\bf ABSTRACT}

\end{center}

It has been shown that the renormalization group (RG) flow
connecting different vacua may occur 
due to the multi-trace operators in four-dimensional CFT. In the AdS/CFT 
set-up, the double trace operators correspond to the tachyonic 
scalar field living in the bulk five-dimensional AdS. In this paper, we
consider the 
five dimensional black hole  as the bulk spacetime instead of the 
pure AdS. In frames of the AdS/CFT correspondence, the AdS black hole 
 is believed to describe dual CFT at finite temperature. 
 From the AdS side, 
we consider the difference of energies between the two vacua by one-loop 
calculation for the scalar field with tachyonic mass in five-dimensional 
AdS black hole. In order to check the dS/CFT correspondence, the corresponding 
calculations are done in five-dimensional deSitter space. 
The difference between the two vacua is specified by the difference of 
the boundary conditions of the scalar field. 
We show that for AdS black hole, there might occur instability which could be 
the manifestation of the Hawking-Page phase transition.
For stable phase of AdS black hole as well as for deSitter bulk, 
proposed c-function found beyond the leading order approximation 
shows the monotonic behaviour consistent with c-theorem. 

\vfill


\newpage

\section{Introduction\label{Sec0}}

The holographic principle realized in string theory in the form of
celebrated AdS/CFT correspondence (for a general review, see \cite{AdS})
brought to our attention not only the new concepts but also better 
understanding of well-known phenomena in high energy physics.
In particulary, well-known RG theory has been enriched by the holographic RG
(for a general introduction and list of references, see \cite{FMS,ACH}.)
In frames of holographic RG it turned out to be easier to study 
the number of questions, like the check of so-called c-theorem 
conjectured in four dimensions in ref.\cite{JLC} by analogy with the 
two-dimensional Zamolodchikov's c-theorem \cite{ABZ}. In fact,
it has been shown \cite{GPPZ2,FGPW} (for earlier studies, see
\cite{ETA,EG}) that four-dimensional c-theorem follows from
classical supergravity side of AdS/CFT. The very important check 
of c-theorem in one-loop approximation (beyond classical supergravity
approximation) has been recently done in ref.\cite{GM}.

The consideration of Gubser-Mitra \cite{GM} is very much related with
Witten's discussion of multi-trace operators \cite{witten2} (see also 
\cite{BSS}) and their dual gauge theory description in (asymptotically) AdS space. 
We may add a term given by the product of the two single trace operators 
${\cal O}$, in the form of ${f \over 2}{\cal O}^2$, to the gauge 
theory Lagrangian. If the coefficient $f$ has the dimension of mass, the 
operator ${f \over 2}{\cal O}^2$ is relevant and the RG flow is generated. 
In the RG flow, the UV fixed point corresponds to $f=0$ and the IR one 
to $f=\infty$. In the AdS side, the two fixed points can be specified by  
the boundary conditions for the massive scalar filed $\phi$ corresponding
to 
the operator ${\cal O}$. In \cite{GM}, the one-loop 
vacuum energy of the scalar field $\phi$ was discussed. The obtained
vacuum energy itself 
diverges but  the difference between the two vacuum energies 
corresponding to the different  boundary conditions of $\phi$ is finite. 
The difference between  two vacuum energies corresponds to the difference
of 
the values of c-functions in the UV and IR fixed points. 
Then we find the conformal invariance is broken by the one-loop effects, which 
lead to non-trivial RG flow. Since the one-loop effect gives the sub-leading 
correction in the $1/N$-expansion, the difference between the values of 
c-function 
in the two vacuum is the sub-leading. 

The present work is motivated by the study done in ref.\cite{GM}.
Using the methods of above paper we search for the description of RG flow 
when double-trace operator for tachyon field is incorporated in the
situation when bulk space is not pure AdS. Specifically,
we start from the tachyon field living in 5d AdS black hole (BH).
(It is not difficult to extend the calculation for other dimensions).
It is known \cite{witten} that AdS/CFT correspondence is applied well 
to situation when bulk space is AdS BH so all discussion of ref.\cite{GM}
may be directly used.
Since the black hole has a temperature, the corresponding dual CFT is 
thermal theory.
It is known that Hawking-Page phase transitions occur between AdS 
BH and global AdS vacuum. These phase transitions have the AdS/CFT 
interpretation \cite{witten} as confinement-deconfinement transitions in dual gauge
theory. Then, it is expected that RG flow induced by double-trace 
operator when bulk space is AdS BH which is dual to thermal QFT should be
qualitatively more
complicated. In particulary, some instability (divergence) corresponding
to the description of phase transition should appear at the one-loop order.
Indeed, such instability is explicitly found in our calculation of 
difference of vacuum energies at IR and UV.

Another interesting aspect of such RG flow investigation is related with
the fact that it may be applied to another duality.
To be specific, we considered also five-dimensional dS bulk space.
This space is expected to be crucial in proposed dS/CFT
correspondence\cite{hull,strominger}. Despite the fact that explicit 
reasonable dual theory in dS$_5$/CFT$_4$ correspondence is not constructed yet,
we found the corresponding RG flow and conjectured c-function  in direct
analogy 
with AdS/CFT. In other words, the check of four-dimensional c-theorem as
it follows from dS/CFT correspondence (for a general introduction, see \cite{SSM})
is done beyond the classical supergravity approximation.

The paper is organized as follows. In the next section we propose
c-function from AdS black hole (no deformation case).
Section three is devoted to the study of massive scalar field in
five-dimensional AdS black hole bulk. Using simplified solution of
Klein-Gordon equation which correctly reproduces the asymptotics 
of the scalar field in the bulk, the corresponding propagator is
constructed. It is used in the calculation of difference of vacuum 
energies for boundary conditions specified by double-trace operator.
The interpretation of instability in such difference as indication to
Hawking-Page phase transition is given. Our investigation permits to
propose c-function in AdS black hole phase using AdS/CFT, while 
usually it is not easy to define c-function in thermal QFT.
Section four is devoted to the study of RG flow found in the similar way
but for dS bulk. This discussion may be relevant for dS/CFT correspondence 
as it gives the information about conjectured c-function  in the situation
when
dual CFT is not found yet. Some outlook is presented in final section.

\section{c-function from AdS black hole: no deformation \label{Sec1}}

Let us start from the discussion of c-function as it may appear from AdS
BH.
In \cite{GPPZ2}, a candidate of the c-function from 5d AdS space has been
proposed in terms of the metric as follows:
\be
\label{gppzC}
c_{\rm GPPZ}=\left({dA \over d z}\right)^{-3}\ ,
\ee
where the metric is taken in the warped form:
\be
\label{gppzC2}
ds^2=dz^2 + \sum_{\mu =1}^4\e^{2A}dx_\mu dx^\mu\ .
\ee
Then for the pure anti-deSitter (AdS) spacetime, where $A={z \over l}$, 
the c-function becomes a constant:
\be
\label{gppzC2b} 
c_{\rm GPPZ}=l^3\ .
\ee 
The c-function (\ref{gppzC}) is positive as it should be.
Note that the c-function is monotonically increasing function of the energy scale $z$.

One may propose a similar c-function for the Schwarzschild-Anti-de 
Sitter (SAdS) space with the flat horizon:
\be
\label{III}
ds^2=g_{\mu\nu}dx^\mu dx^\nu = - \e^{2\rho}dt^2 
+ \e^{-2\rho}dr^2 + r^2 \sum_{i=1,2,3}\left(dx^i\right)^2\ ,
\quad \e^{2\rho}={r^2 \over l^2} - {\mu \over r^2}\ .
\ee
If we redefine the coordinates $r$ and $x^i$ by new ones 
$\eta$ and $\hat x^i$ as
\be
\label{I}
r^2 = l\mu^{1 \over 2}\cosh {2\eta \over l}\ ,\quad 
x^i=l^{-{1 \over 2}}\mu^{-{1 \over 4}}\hat x\ ,
\ee
and furthermore  Wick-rotate the time coordinate $t$ to $\theta$ by 
\be
\label{II}
t={i l^{3 \over 2} \over 2\mu^{1 \over 4}}\theta\ ,
\ee
the metric  (\ref{III}) can be rewritten as 
\be
\label{IV}
ds^2 = {l^2 \sinh^2 {2 \eta \over l} \over 4\cosh {2\eta \over l}}d\theta^2 
+ d\eta^2 + l^2\cosh {2\eta \over l}\sum_{i=1,2,3}\left(d\hat x^i\right)^2\ .
\ee
A little bit surprizingly, the obtained metric does not depend on $\mu$, 
which parametrizes the black hole (BH). As is clear from (\ref{I}), the
pure 
AdS ($\mu\to 0$) corresponds to $\eta\to \infty$ limit and the large black hole 
 corresponds to $\eta\to 0$ limit. The horizon exists at $\eta=0$ or 
$r^2 = l\mu^{1 \over 2}$. In order to avoid the orbifold singularity at the 
horizon, the Euclid time coordinate $\theta$ has a period of $2\pi$. 

The obtained metric (\ref{IV}) does not have the warped form as in 
(\ref{gppzC2}). Since $\sqrt{g}=\e^{4A}$ in (\ref{gppzC2}),  identifying 
$\eta$ with $z$ in (\ref{gppzC2}), one may propose a c-function $c_1$ as
\be
\label{c1}
c_1 = \left({1 \over 8}{d\ln g \over d\eta}\right)^{-3}\ .
\ee
When the metric can be written in the warped form (\ref{gppzC2}), 
 identifying $\eta$ with $z$, (\ref{c1}) reproduces (\ref{gppzC}). When 
the metric 
has a warped form (\ref{gppzC2}), the scale transformation $x^\mu\to \e^\lambda x^\mu$ 
with a constant parameter $\lambda$ of the transformation can be absorbed into the shift 
of $A$ like $A\to A - \lambda$. If we identify the coordinate $z$ as a
parameter of the 
scale, the scale invariance requires ${d^2 A \over dz^2}=0$. Similarly for the metric 
(\ref{IV}), the scale transformation $\theta\to \e^\lambda \theta$ and 
$x^i\to \e^\lambda x^i$ with a constant $\lambda$ can be absorbed into the scale 
transformation of the metric ${l^2 \sinh^2 {2 \eta \over l} \over 4\cosh {2\eta \over l}}
\to \e^{-2\lambda}{l^2 \sinh^2 {2 \eta \over l} \over 4\cosh {2\eta \over l}}$ and 
$l^2\cosh {2\eta \over l}\to \e^{-2\lambda}l^2\cosh {2\eta \over l}$. Then if $\eta$ 
corresponds to the parameter of the scale transformation,
 the scale invariance  requires 
${d^2 \over d\eta^2}\left( {l^2 \sinh^2 {2 \eta \over l} \over 4\cosh {2\eta \over l}}
\right) = {d^2 \over d\eta^2}\left(l^2\cosh {2\eta \over l}\right)=0$.
 Then one may define 
two kinds of c-function corresponding to $(\theta,\theta)$- and $(i,j)$-components of 
the metric. The c-function (\ref{c1}) is a kind of the average of the two 
c-functions. At least, at the RG fixed point, if exists, where there is a scale invariant, 
we find ${dc_1 \over d\eta}=0$.  

By using (\ref{IV}), one gets
\be
\label{c2}
g= {l^8 \sinh^2 {2 \eta \over l}\cosh {2\eta \over l} \over 4}
= {l^8 \sinh^2 {4 \eta \over l} \over 16}\ .
\ee
Then 
\be
\label{c3}
c_1 = l^3 \tanh^3 {4 \eta \over l} 
\ee
This expression reproduces (\ref{gppzC2b}) in the pure AdS limit 
($\eta\to\infty$). We also note that $c_1$ is monotonically increasing 
function of $\eta$ and therefore the energy scale. It is interesting 
that  c-function vanishes at the horizon $\eta=0$, which can 
be identified with the large black hole limit. 
The central charge, in general, measures the massless degrees of freedom.
Then the vanishing central charge means that there are 
no massless modes. At the horizon, the time coordinate becomes degenerate 
(as well-known, anything near the horizon moves slowly for the observer far from 
the horizon). Then effectively, any mass or energy goes to infinity and 
only constant modes can survive there. 

\begin{figure}[htbp]
\begin{center}
\unitlength=0.8mm
\begin{picture}(160,100)
\thicklines
\put(20,20){\vector(1,0){120}}
\put(20,20){\vector(0,1){60}}
\put(141,18){${\eta \over l}$}
\put(10,82){c-function(s)}
\put(18,15){$0$}
\qbezier[100](20,60)(80,60)(140,60)
\put(110,62){$c_{\rm GPPZ}$}
\qbezier(20,20)(60,20)(80,40)
\qbezier(80,40)(99,59)(140,59)
\put(80,35){$c_1$}
\put(15,60){$l^3$}
\end{picture}
\end{center}
\caption{ $c_{\rm GPPZ}$  
(\ref{gppzC2b}) for the pure AdS is given by the dotted line and 
$c_1$ (\ref{c3}) by the solid line. 
\label{Fig0}}
\end{figure}
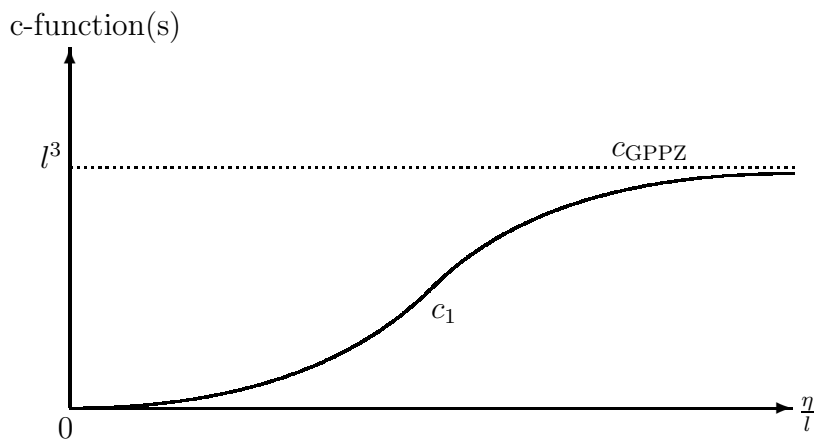

\section{The scalar field in AdS black hole and one-loop vacuum energy for
tachyon \label{Sec1b}}

In AdS/CFT correspondence, RG flow can be generated by the double trace operator.
On supergravity side, the double trace operator corresponds to the scalar 
field with tachyonic mass.   
We now consider the propagation of the free scalar field $\phi$ 
with tachyonic mass $m$ ($m^2<0$):
\be
\label{T1}
S_\phi = \int d^5x \sqrt{-g_{(5)}}\left\{-{1 \over 2}\partial_\mu \phi 
\partial^\mu \phi - {m^2 \over 2}\phi^2\right\}
\ee
in five-dimensional AdS BH.
In the black hole spacetime (\ref{IV}), which has been Wick-rotated 
into the Euclidean signature, the Klein-Gordon equation has the following form: 
\bea
\label{V}
&&{4\cosh {2\eta \over l} \over l^2 \sinh^2 {2\eta \over l}}\partial_\theta^2 \phi
+ {1 \over \sinh{2\eta \over l} \cosh{2\eta \over l}}\partial_\eta
\left(\sinh{2\eta \over l} \cosh{2\eta \over l}\partial_\eta \phi\right) \nn
&& + {1 \over \cosh {2\eta \over l}}\sum_{i=1,2,3}{\partial^2 \phi \over 
\partial\tilde {x^i}^2} - m^2 \phi = 0\ .
\eea
Taking the plane wave, one replaces 
$\partial_\theta^2$ and ${\partial^2 \over \partial\tilde {x^i}^2}$ by  
\be
\label{VI}
\partial_\theta^2\to - n^2\ ,\quad 
\sum_{i=1,2,3}{\partial^2 \over \partial\tilde {x^i}^2}\to -k^2\ .
\ee
Here $n$ is an integer. With redefined  scalar filed $\phi$ 
\be
\label{VII}
\phi={\tilde \phi \over \sinh^{1 \over 2}{2\eta \over l} 
\cosh^{1 \over 2}{2\eta \over l}}\ ,
\ee
 the Schr\"odinger-like equation looks like:
\be
\label{VIII}
-\partial_\eta^2 \tilde \phi + \left\{ -{1 \over l^2}\left({1 \over 
\sinh^2 {2\eta \over l}\cosh^2 {2\eta \over l}} - 4\right) 
+ {4n^2 \cosh {2\eta \over l} \over l^2 \sinh^2 {2\eta \over l}} 
+ {k^2 \over \cosh^2 {2 \eta \over l}}\right\}\tilde\phi = m^2 \tilde \phi\ .
\ee
When $\eta$ is small, Eq.(\ref{VIII}) behaves as
\be
\label{IX}
-\partial_\eta^2 \tilde \phi + {1 \over \eta^2}\left\{ -{1 \over 4} + n^2\right\}\tilde\phi 
= 0\ .
\ee
Then 
\be
\label{X}
\tilde\phi \sim \eta^{{1 \over 2} \pm n}\ ,
\ee
or by using (\ref{VII}), we find 
\be
\label{XI}
\phi\sim \eta^{\pm n}\ .
\ee
On the other hand, when $\eta$ is large, Eq.(\ref{VIII}) behaves as 
\be
\label{XII}
-\partial_\eta^2 \tilde \phi +{4 \over l^2} \phi = m^2 \tilde \phi\ ,
\ee
and one gets
\be
\label{XIII}
\tilde\phi \sim \e^{\pm \eta\sqrt{{4 \over l^2} + m^2}}\ ,
\ee
or 
\be
\label{XIV}
\phi\sim \e^{\eta\left(-{2 \over l} \pm \sqrt{{4 \over l^2}+m^2}\right)}\ .
\ee
Using the coordinate $r$, we have 
\be
\label{XIVb}
\phi\sim r^{-\Delta_\pm}\ ,\quad \Delta_\pm \equiv 2 \pm \sqrt{4 + m^2 l^2}\ ,
\ee
which reproduces the usual AdS/CFT results, that is, the power law behaviour corresponds 
to the operator in CFT with conformal weight $\Delta_\pm$.

Since it is difficult to solve the Klein-Gordon equation exactly in the Schwarzschild-AdS 
background, we now consider a simplified model. From Eqs.(\ref{IX}) and (\ref{XII}), we can 
find that the $\tilde x^i$ or $k$-dependence is not essential for both of small $\eta$ and 
large $\eta$ cases. Then one may neglect the $k$-dependence for the qualitative argument. 
We now approximate the Klein-Gordon equation (\ref{VIII}) by 
\be
\label{XV}
\partial_\eta^2 \tilde\phi + \left\{ - {4 \over l^2}
+ \left( {1 \over 4} - n^2 \right){1 \over \eta^2}\right\}\tilde\phi 
= m^2 \tilde \phi\ , 
\ee
which reproduces (\ref{IX}) for small $\eta$ and (\ref{XII}) for large $\eta$. 
We also assume, instead of (\ref{VII}), 
\be
\label{XVI}
\phi=\sqrt{l \over 2\eta}\e^{-{2\eta \over l}}\tilde\phi\ ,
\ee
which coincides with (\ref{VII}) in the large $\eta$ (except $\eta^{-{1 \over 2}}$ 
and small $\eta$ limits.
The solution of (\ref{XV}) is given by the modified Bessel functions $I_n$ and 
$K_n$:
\be
\label{XVIb}
\tilde \phi =\eta^{1 \over 2}\left(\alpha I_n\left(\gamma\eta\right) 
+ \beta K_n\left(\gamma\eta\right)\right)\ .
\ee
Here
\be
\label{XVII}
\gamma\equiv\sqrt{{4 \over l^2} + m^2}\ .
\ee
Then one gets
\be
\label{XVIc}
\phi =\sqrt{l \over 2}\e^{-{2\eta \over l}}\left(\alpha I_n\left(\gamma\eta\right) 
+ \beta K_n\left(\gamma\eta\right)\right)\ .
\ee
Since $I_n(z)\sim {\e^z \over \sqrt{2\pi z}}$ and $K_n (z)\sim \e^{-z} \sqrt{\pi 
\over 2z}$ for large $z$, when $\eta$ is large, $\phi$ given by combining 
(\ref{XVI}) and (\ref{XVIb}) behaves as 
\be
\label{XVIII}
\phi \sim \alpha {\e^{ \left(- {2 \over l} + \sqrt{{4 \over l^2} + m^2}\right)\eta} 
\over \sqrt{2\pi}} + \beta \e^{ \left(- {2 \over l} - \sqrt{{4 \over l^2} + m^2}\right)\eta} 
\sqrt{2 \over \pi}\ .
\ee
In \cite{GM}, the deformation of the CFT by the double trace operator 
\be
\label{XIX}
{f \over 2}\int d^4 x {\cal O}^2
\ee
has been discussed from the AdS side. The parameter $f$ corresponds to the 
boundary condition for the bulk scalar field. In the present case, the boundary 
condition looks like
\be
\label{XX}
f={\alpha \over \pi\beta}\ .
\ee
With two independent solutions for $\tilde \phi$  
\be
\label{XXI}
\tilde \phi_a =\eta^{1 \over 2}\left(\alpha_a I_n\left(\gamma\eta\right) 
+ \beta_a K_n\left(\gamma\eta\right)\right)\ ,\quad a=1,2\ ,
\ee
the propagator $\tilde G(\eta,\xi)$ of $\tilde \phi$ can be constructed as 
\be
\label{XXII}
\tilde G(\eta, \xi)= {\phi_1 (\eta) \phi_2(\xi) \theta(\eta - \xi) 
+ \phi_1 (\xi) \phi_2(\eta) \theta(\xi - \eta) 
\over \gamma\left(\alpha_1 \beta_2 - \beta_1 \alpha_2\right)}\ .
\ee
In fact $\tilde G(\eta, \xi)$ satisfies the following equation, 
corresponding to 
(\ref{XV}):
\be
\label{XXIII}
-\partial_\eta^2 \tilde G(\eta, \xi) + \left\{ - {4 \over l^2} - m^2 
+ \left({ 1\over 4} - n^2 \right){1 \over \eta^2}\right\}\tilde G(\eta, \xi) = 
\delta(\eta - \xi)\ .
\ee
Then from (\ref{XVI}) the propagator $G(\eta, \xi)$ for $\phi$ is given by
\be
\label{XXIV}
G(\eta, \xi; m^2)= \sqrt{l \over 4\eta\xi}\e^{-{2\left(\eta+\xi\right) \over l}}
\tilde G(\eta, \xi)\ .
\ee
Then the vacuum energy  is expressed as \cite{GM}
\bea
\label{XXV}
V(\eta)&=&-{1 \over 2}\int_{-{4 \over l^2}}^{m^2} d\tilde m^2 G(\eta,\eta; \tilde m^2) \nn 
&=&- {1 \over 2}\int_0^{\gamma^2} d\tilde \gamma^2 G(\eta,\eta; \tilde m^2(\tilde\gamma) ) \nn 
&=& -{1 \over 2}\int_0^{\gamma^2} d\tilde \gamma^2 {l \e^{-{4\eta \over l}} \over 
2\tilde\gamma \sin\left(\varphi_1 - \varphi_2\right)} \sum_{n=-\infty}^\infty\left\{
\cos\varphi_1 \cos \varphi_2 I_n(\tilde\gamma\eta)^2 \right. \nn
&& \left. + \sin\left(\varphi_1 + \varphi_2\right) I_n(\tilde\gamma\eta) K_n(\tilde\gamma\eta) 
+ \sin\varphi_1 \sin \varphi_2 K_n(\tilde\gamma\eta)^2 \right\}\ .
\eea
Here we denote 
\be
\label{XXVI}
\alpha_a = \cos\varphi_a\ ,\quad \beta_a=\sin\varphi_a\ ,
\ee
since the overall factor of $\alpha_a$ and $\beta_a$ is irrelevant for the 
propagator. By comparing (\ref{XX}), (\ref{XXII}) and (\ref{XXVI}), we find 
\be
\label{XXVII}
f={1 \over \pi\tan\varphi_1}\ .
\ee
On the other hand, the parameter $\varphi_2$ could be determined from the 
boundary condition at the horizon $\eta=0$. It is natural that the 
scalar field is regular at the horizon. Then  
\be
\label{XXVIIb}
\varphi_2=0\ .
\ee
Since
\bea
\label{XXVIII}
\sum_{n=-\infty}^\infty I_n(z)^2= I_0(2z)\ ,\nn
\sum_{n=-\infty}^\infty I_n(z) K_n(z)= K_0(2z)\ ,
\eea
one gets 
\be
\label{XXIX}
V(\eta)= -{1 \over 2}\int_0^{\gamma^2} d\tilde\gamma^2 {l \e^{-{4\eta \over l}} \over 
2\tilde\gamma\sin\varphi_1 } \left\{\cos\varphi_1 I_0(2\tilde\gamma\eta) + \sin\varphi_1 
K_0(2\tilde\gamma\eta) \right\}\ .
\ee
The remark is in order. The obtained vacuum energy is generally divergent
quantity. The calculation of vacuum energy as finite quantity requires 
the application of some regularization. Using mainly zeta-regularization,
the calculation of above vacuum energy for bulk scalars and spinors 
on five-dimensional AdS space was performed in refs.\cite{GPT,NOZ,
HKP,FMT,NS,SS}. Instead of repeating such complicated calculations,
we follow to the prescription developed in \cite{GM} and calculate the
difference between vacuum energies corresponding to different boundary 
conditions specified by the double-trace operator.
For pure AdS space such difference corresponds to difference between IR
and UV points (end-points of RG flow) and is finite.

Then  the difference of vacuum energies specified by the two boundary
conditions corresponding 
to $\varphi_1=\Theta_{(1)}$ and $\varphi_1=\Theta_{(2)}$ is given by
\bea
\label{XXX}
\delta V&=&V\left(\eta,\varphi_1=\Theta_{(1)}\right) 
 - V\left(\eta,\varphi_1=\Theta_{(2)}\right) \nn
&=& -\left(\cot \Theta_{(1)} - \cot \Theta_{(2)} \right)
{1 \over 2}\int_0^\gamma d\gamma \e^{-{4\eta \over l}} I_0(2\gamma\eta) \nn
&=& -\left(\cot \Theta_{(1)} - \cot \Theta_{(2)} \right)
{\e^{-{4\eta \over l}} \gamma \over 2}{}_1 F_2\left({1 \over 2};1,{3 \over 2}; \gamma^2\eta^2
\right) \ .
\eea
Here ${}_1 F_2\left(a;b,c;z\right)$ is a hypergeometric function. 
When $\gamma=0$, $\delta V=0$. This is consistent with the corresponding CFT. 
Since  $\gamma=0$ corresponds to $\Delta_-=2$, the deformation of the CFT 
is marginal and the central charge does not change, that is, $\delta V=0$. 
Since   $f\to \infty$ corresponds to $\varphi\to 0$ 
and $f\to 0$ to $\varphi\to {\pi \over 2}$ and , if we consider the difference between 
$f\to \infty$ and $f\to 0$, that is $\Theta_1\to 0$ and 
$\Theta_2\to {\pi \over 2}$, $\delta V$ diverges. This is different from
the result in 
pure AdS case and is presumably related with phase transitions as we
argue below. For the pure AdS case, one gets\cite{GM} 
\be
\label{XXXI}
\delta V={1 \over 12\pi^2 l^5} \left\{{\left(\Delta_- -2\right)^3 \over 3} 
 - {\left(\Delta_- -2\right)^5 \over 5} \right\}
=-{1 \over 12\pi^2 l^5} \left\{{\gamma^3l^3 \over 3} 
 - {\gamma^5 l^5 \over 5} \right\}\ .
\ee
When $\eta$ is large, since $I_0(z)\sim {\e^z \over \sqrt{2\pi z}}$, 
we find $\delta V$  (\ref{XXX}) behaves as 
\be
\label{XXXII}
\delta V\sim -{\left(\cot \Theta_{(1)} - \cot \Theta_{(2)} \right) \over 4}
\int_0^\gamma d\gamma {\e^{-\left({4 \over l} - 2\gamma\right)\eta} 
\over \sqrt{\gamma\pi \eta}}\ .
\ee
Therefore if $l\gamma < 2$ ($l\gamma > 2$), $\delta V$ exponentially decreases 
(increases). From the viewpoint of CFT, $l\gamma = 2$ corresponds to 
$\Delta_-=0$. 

\begin{figure}%[htbp]
\begin{center}
\epsffile{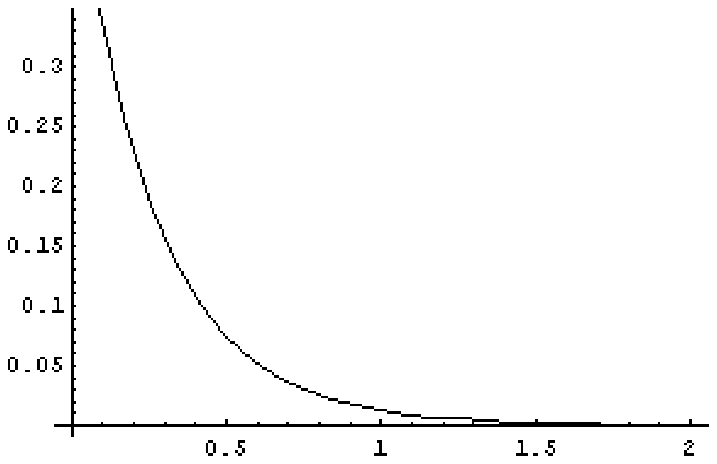}
\end{center}
\caption{\label{Fig1}
$F(\gamma;\eta)$ versus ${\eta \over l}$ for $l\gamma=1$. $F(\gamma;\eta)$ 
decreases exponentially.}
\end{figure}

\begin{figure}%[htbp]
\begin{center}
\epsffile{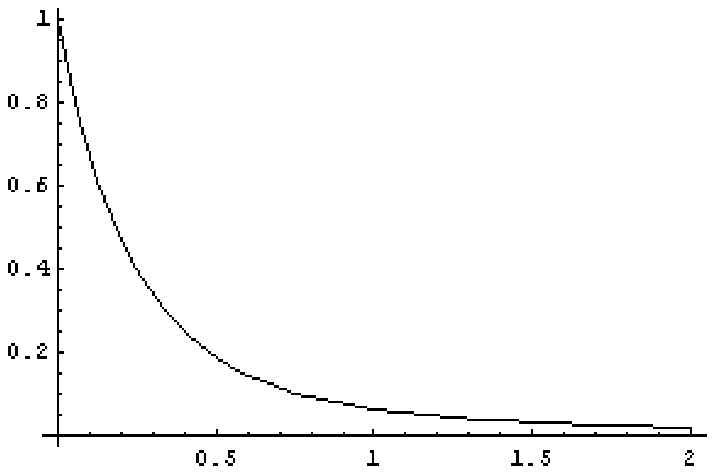}
\end{center}
\caption{\label{Fig2}
$F(\gamma;\eta)$ versus ${\eta \over l}$ for $l\gamma=2$. $F(\gamma;\eta)$ 
decreases exponentially.}
\end{figure}

\begin{figure}%[htbp]
\begin{center}
\epsffile{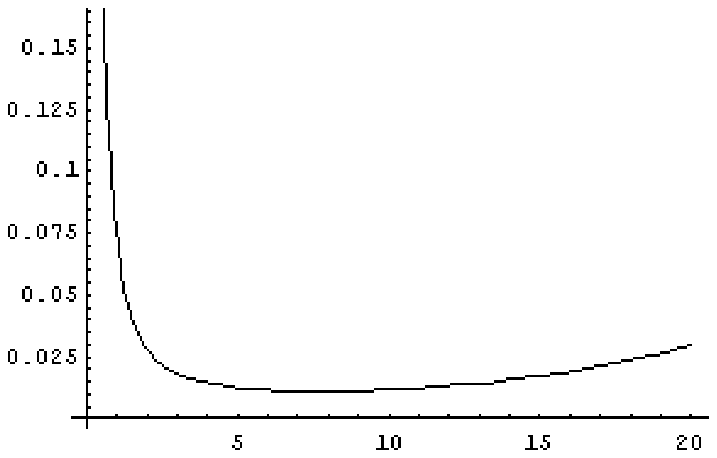}
\end{center}
\caption{\label{Fig21}
$F(\gamma;\eta)$ versus ${\eta \over l}$ for $l\gamma=2.1$. $F(\gamma;\eta)$ 
increases exponentially for large ${\eta \over l}$ and there appears a
minimum.}
\end{figure}

Let us define $F(\gamma;\eta)$ by
\be
\label{XXXIII}
F(\gamma;\eta)\equiv
{1 \over 2}\int_0^\gamma d\gamma \e^{-{4\eta \over l}} I_0(2\gamma\eta) 
={\e^{-{4\eta \over l}} \gamma \over 2}{}_1 F_2\left({1 \over 2};1,{3 \over 2}; \gamma^2\eta^2
\right) \ ,
\ee
which appears in (\ref{XXX}). Some plots have been given in Figure \ref{Fig1} for 
$l\gamma=1$, Figure \ref{Fig2} for $l\gamma=2$, and Figure \ref{Fig21} for 
$l\gamma=2.1$ as functions of ${\eta \over l}$, which corresponds to the 
horizontal axis. When $l\gamma\leq 2$, $F(\gamma;\eta)$ is a monotonically 
decreasing function of ${\eta \over l}$. When $l\gamma>2$, $F(\gamma;\eta)$ 
increases exponentially for large ${\eta \over l}$ and there appears a
minimum. 
The minimum appears at ${\eta \over l}=7.6068\cdots$ and the minimal value 
of $F$ is $0.0108396\cdots$. When $l\gamma=4$, there appears a minimum at 
${\eta \over l}=0.4176\cdots$ and the minimum value of $F$ is $0.910144\cdots$. 
In the language of the holographic renormalization group, the coordinate 
$\eta$ corresponds to the parameter of the scaling. If $\delta V$ corresponds to 
the difference of the central charges, the minimum, which appears for $l\gamma>2$, 
should correspond to the renormalization fixed point. On the other hand, as 
explained before, large $\eta$ corresponds to the pure AdS and small $\eta$ to 
the large AdS black hole. Then the minimum might divide two phases
corresponding to 
the pure AdS or small black hole phase and the large black hole phase. 
Since the Hawking temperature is given by $T_{H} = {r_H \over \pi l^2}$, 
the larger black hole has the higher temperature. Here $r_H$ is the radius of the 
horizon. Then the phase transition, if exists, 
might correspond to the thermal transition of the CFT. 

Indeed, it has been suggested by Hawking and Page \cite{HP}, there 
is a phase transition between AdS BH spacetime 
and global AdS vacuum. BH is stable at high 
temperature but it becomes unstable at low temperature, which can be 
understood from the free energy $F$:
\be
\label{F1}
F= -{V_{3} \over \kappa^2}r_{H}^{2} \left( {r_{H}^{2} \over l^{2}}
 - {k \over 2} \right)
\ee 
Here if $k>0$ the boundary can be three dimensional 
sphere, if $k<0$, hyperboloid, or if $k=0$, flat space. 
Properly normalizing the coordinates, one can choose $k=2$, $0$, 
or $-2$. Then if $k=2$, the free energy $F$ vanishes at $r_H=l$, 
which is the critical point of the phase transition. From the point 
of view of the AdS/CFT correspondence \cite{AdS}, 
this phase transition  corresponds to the 
confinement-deconfinement transition in dual gauge theory 
\cite{witten}. 
Eq.(\ref{F1}) seems to tell that there is no phase transition when 
$k=0$ or $k=-2$, which is not always true. There is an argument 
that it is easier to subtract the action of AdS soliton \cite{HM} 
instead of the vacuum AdS \cite{SSW}. In this case, 
besides the temperature, the area of the horizon becomes an independent 
parameter on which the thermodynamical quantities depend. 

When $k=0$, the metric of SAdS BH  (\ref{III}) can be written as 
\bea
\label{k0bh}
ds_{\rm BH}^2 &=& - \e^{2\rho_{\rm BH}(r)} dt_{\rm BH}^2 
+ \e^{-2\rho_{\rm BH}(r)}dr^2
 + r^2\left(d\phi_{\rm BH}^2 + \sum_{i=1,2}
\left(dx^i\right)^2\right)\ , \nn
\e^{2\rho_{\rm BH}(r)}&=&{1 \over r^2}\left\{-\mu_{\rm BH} 
+ {r^4 \over l^2}\right\}\ .
\eea
In (\ref{k0bh}), we choose a torus for the $k=0$ Einstein 
manifold for simplicity. The coordinates of the torus are 
$\phi_{\rm BH}$ and $\left\{x^1,x^2\right\}$. One assumes 
$\phi_{\rm BH}$ has a period of $\eta_{\rm BH}$:
$\phi_{\rm BH}\sim \phi_{\rm BH} + \eta_{\rm BH}$.
The AdS soliton solution can be obtained by exchanging the 
signature of $t_{\rm BH}$ and $\phi_{\rm BH}$ as 
$t_{\rm BH}\rightarrow i\phi_s$ and 
$\phi_{\rm BH}\rightarrow it_s$.
Then the metric of the AdS soliton is given by
\bea
\label{k0sltn}
ds_s^2 &=& - r^2 dt_s + \e^{-2\rho_s(r)}dr^2
 + \e^{2\rho_s(r)} d\phi_s^2 + r^2\sum_{i=1,2}
\left(dx^i\right)^2\ , \nn
\e^{2\rho_s(r)}&=&{1 \over r^2}\left\{-\mu_s 
+ {r^4 \over l^2}\right\}\ .
\eea
The free energy $F$ is obtained as follows:
\be
\label{F}
F= -{\eta_{\rm BH} V_2 l^6 \over \kappa^2}\left\{\left(\pi T_{H}\right)^4 
 - \left({\pi \over l\eta_{\rm BH}}\right)^4 \right\} \ .
\ee
Eq.(\ref{F}) tells that there is a phase transition at
\be
\label{Tc}
T_{\rm BH}={1 \over l\eta_{\rm BH}}\ .
\ee
Hence,
when $T_{H}>{1 \over \eta_{\rm BH}}$, the black 
hole is stable but when $T_{H}<{1 \over \eta_{\rm BH}}$, 
the black hole becomes unstable and the AdS soliton  is 
preferred. Although the AdS soliton solution is not connected with the
SAdS 
solution by any continious parameter, the maximum of $\delta V$
(\ref{XXX}), 
which appears when $l\gamma > 2$, might correspond to the above phase 
transition.  

Note that if one defines a c-function by
\bea
\label{CI}
c_2&=&\delta V - \delta V(\eta=0) \nn
&=& -\left(\cot \Theta_{(1)} - \cot \Theta_{(2)} \right)
\left(F(\gamma;\eta) - F(\gamma;\eta=0)\right)\ ,
\eea
the c-function $c_2$ is a monotonically increasing function of $\eta$ 
if $0<l\gamma\leq 2$ and $\cot \Theta_{(1)} - \cot \Theta_{(2)}> 0$.  
We also note that $c_2$ vanishes by construction, at $\eta=0$ or 
large black hole limit as $c_1$ (\ref{c3}). 

We now consider the difference between two black hole solutions. 
Let a radial coordinate of one of the black hole solutions is associated
with a
mass parameter $\mu_1$ as $\eta_1$ and another with $\mu_2$ as $\eta_2$. 
Then Eq.(\ref{I}) shows
\be
\label{XXXIV}
\xi\equiv \left({\mu_1 \over \mu_2}\right)^{-{1 \over 2}}={\cosh {2\eta_1 \over l} 
\over \cosh {2\eta_2 \over l}}\ ,
\ee
We now assume $\mu_2>\mu_1$ and therefore $\eta_1>\eta_2$ and $\xi>1$. 
In the UV region, $\eta_{1,2}$ becomes large
\be
\label{XXXV}
\xi=\left({\mu_1 \over \mu_2}\right)^{-{1 \over 2}}\to \e^{2\left(\eta_1 
 -\eta_2\right) \over l}\ .
\ee
As an IR region, we consider a limit $\xi_2\to 0$. Then in the limit
\be
\label{XXXVI}
\e^{2\eta_1 \over l}\sim \xi + \sqrt{\xi^2 - 1}\ .
\ee
Here the sign in front of $\sqrt{\xi^2 - 1}$ can be fixed since 
$\eta_1>\eta_2=0$ $\e^{2\eta_1 \over l}>1$. 

When $\eta_1$ and $\eta_2$ are large,  using (\ref{XXXV}) one gets
\bea
\label{XXXVII}
F(\gamma;\eta_1)&\sim& {1 \over 4\sqrt{\pi}}\int_0^\gamma d\tilde \gamma 
{\e^{-{4\eta_1 \over l}+2\tilde\gamma\eta_1} \over \sqrt{\tilde \gamma 
\eta_1}} \nn
&=& {1 \over 4\sqrt{\pi}}\sqrt{\eta_2 \over \eta_1}
\e^{-{4\left(\eta_1 -\eta_2 \right)\over l}}
\int_0^\gamma 
d\tilde \gamma {\e^{-{4\eta_2 \over l}+2\tilde\gamma\eta_2 + 
{\tilde\gamma \over l}\ln\xi } \over \sqrt{\tilde \gamma \eta_1}} \ ,
\eea
and 
\be
\label{XXXVIII}
\e^{4\eta_1 \over l}F(\gamma;\eta_2) < \e^{4\eta_2 \over l}F(\gamma;\eta_1)
< \e^{4\eta_1 \over l}F(\gamma;\eta_2)\e^{{\gamma \over l}\ln\xi}\ ,
\ee
since ${\eta_1 \over \eta_2}\to 1$. 
Let denote $\delta V$ (\ref{XXX}) or (\ref{XXXII}) corresponding to 
$\eta_1$ as $\delta V_1$ and that to $\eta_2$ as $\delta V_2$. Then if 
$(\cot \Theta_{(1)} - \cot \Theta_{(2)}) >0$,
Eq.(\ref{XXXVIII}) indicates 
that 
\be
\label{XXXIX}
\e^{-{4\over l}\ln\xi}\delta V_2 < \delta V_1 < \delta V_2 
\e^{-{4 -2\gamma \over l}\ln\xi}\ ,
\ee
in the limit that $\eta_1$ and $\eta_2$ become large by keeping the difference 
finite : $\eta_1 - \eta_2 = {l \over 2}\ln \xi$. On the other hand, in the 
limit $\eta_2\to 0$, one has
\be
\label{XXXX}
\delta V_2 - \delta V_1 = 
-\left(\cot \Theta_{(1)} - \cot \Theta_{(2)} \right)
\left\{{\gamma \over 2} - F(\gamma;\eta_1)\right\}\ .
\ee
Here $\eta_1$ is given by (\ref{XXXVI}). The right hand side of (\ref{XXXX}) 
 vanishes when $\eta_1=0$. When $\gamma\leq 2$, since $F(\gamma;\eta_1)$ 
is monotonically decreasing function as in Figures \ref{Fig1} and \ref{Fig2}, 
we have if $(\cot \Theta_{(1)} - \cot \Theta_{(2)}) >0$
\be
\label{XXXXI}
\delta V_2 < \delta V_1 \ .
\ee
Even if $\gamma > 2$, $F(\gamma;\eta_1)$ is monotonically decreasing function
for small $\eta_1$ as in Figure \ref{Fig21} and we obtain (\ref{XXXXI}) but if 
$\eta_1$ becomes larger, we have 
\be
\label{XXXXII}
\delta V_2 > \delta V_1 \ .
\ee
 Eqs.(\ref{XXXXI}) and (\ref{XXXXII}) indicate that there might be 
a phase transition ( the Hawking-Page phase transition) if the black hole
radius 
becomes large. 
The point where $\delta V_2 = \delta V_1$, of course, depends on the value of 
$\eta_{1,2}$. The point can be read off from Figure \ref{Fig21} by
shifting $\eta$. 
The line of the graph obtained by the shift always crosses the line 
of the original graph. The crossing point corresponds to $\delta V_2 = \delta V_1$. 
Thus, we demonstrated that away from phase transition , in AdS BH phase
the conjectured c-function\footnote{It is known that formulation of RG in
thermal QFT 
is ambigious as not only high energy but also high temperature 
limit may be considered. Then, the construction of c-function in
thermal QFT is not so trivial. It seems that AdS/CFT
correspondence as discussed above for AdS BH may provide an idea about
correct 
c-function in dual thermal field theory.}
deformed by double-trace operator shows the monotonically
increasing behaviour in accord with c-theorem.
Note also that above consideration may be generalized to other odd
dimensions. 

The confinement-deconfinement transition at finite 
temperature occurs even without the double trace operator. 
Using the propagator (\ref{XXIV}) of the tachyonic scalar filed $\phi$, 
we may consider the correlation function of the operators ${\cal O}$ 
corresponding to $\phi$. From the AdS/CFT correspondence, the correlation 
function of two ${\cal O}$ is proportional to the 
propagator:
\be
\label{ppp1}
\left< {\cal O}{\cal O} \right>\propto \left.G\left(\eta,\eta;m^2\right)
\right|_{\eta\to\, {\rm boundary}}\ .
\ee
The connected part of the two double trace operators is 
\be
\label{ppp2}
\left< {f \over 2}{\cal O}^2 {f \over 2}{\cal O}^2 \right>
\propto \left(\left.G\left(\eta,\eta;m^2\right)
\right|_{\eta\to\, {\rm boundary}}\right)^2\ .
\ee
We now consider the correlation functions in the coordinate space and use the 
Euclidean time $\tau$, which is defined by
\be
\label{ppp3}
\tau ={l^{3 \over 2} \over 2\mu^{1 \over 4}}\theta={\theta \over 2\pi T_H}\ ,
\ee
as in (\ref{II}). Then by summing up the Fourier expansion, we obtain
\bea
\label{ppp4}
G_4\left(\eta,\tau\right)&=&\sum_{n=-\infty}^\infty G\left(\eta,\eta;m^2\right)
\e^{i2\pi n T_H \tau} \nn
&=& \sum_{n=-\infty}^\infty \e^{i2\pi n T_H \tau}
{l\e^{- {4\eta \over l}} \over 2\gamma \sin\varphi_1}\left\{\cos\varphi_1 
I_n\left( \gamma\eta\right)^2 + \sin\varphi_1 I_n\left(\gamma\eta\right) 
K_n\left(\gamma\eta\right)\right\} \nn
&=& {l\e^{- {4\eta \over l}} \over 2\gamma \sin\varphi_1}\left\{\cos\varphi_1 
I_0\left( \gamma\eta\sqrt{2\left(1 + \cos\left( 2\pi T_H \tau\right)\right)}\right) \right.\nn
&& \left. + \sin\varphi_1 K_0\left(\gamma\eta\sqrt{2\left(1 
+ \cos\left( 2\pi T_H \tau\right)\right)}\right)\right\} \ .
\eea
Here $\varphi_2=0$ is assumed as in (\ref{XXVIIb}) and  the
following formulae are used:
\bea
\label{ppp5}
&& \sum_{n=-\infty}^\infty \sin n\theta\, I_n(w) I_n(z) = 0\ ,\nn 
&& \sum_{n=-\infty}^\infty \cos n\theta\, I_n(w) I_n(z) = I_0\left(\sqrt{w^2 + z^2 
+ 2zw \cos\theta}\right)\ ,\nn
&& \sum_{n=-\infty}^\infty \sin n\, \theta I_n(w) K_n(z) = 0\ ,\nn 
&& \sum_{n=-\infty}^\infty \cos n\theta\, I_n(w) K_n(z) = K_0\left(\sqrt{w^2 + z^2 
+ 2zw \cos\theta}\right)\ .
\eea
As 
\be
\label{ppp6}
r^2 = \pi^2 l^4 T_H \cosh{2 \eta \over l}\ ,
\ee
from (\ref{I}), at low temperature limit $T_H\to 0$  keeping 
$\tau$ and $r$ to be finite, we find
\be
\label{ppp7}
G_4\left(\eta,\tau\right)\sim {l\e^{- {4\eta \over l}} \over 2\gamma \sin\varphi_1}
\left\{\cos\varphi_1 {2\gamma\eta \over \sqrt{4\pi \gamma\eta}} 
+ \sin\varphi_1  \ln \left(2\pi\gamma\eta T_H \tau\right)\right\} \ .
\ee
The last logarithmic term depending on $\tau$ is similar to the Green function 
for the massless scalar in two dimensions, which may occur since we
neglected the 
$k$-dependence and substantially reduced the spacetime dimensions. 
On the other hand, at high temperature limit, which could correspond
to 
the limit of $\eta\to 0$, we find 
\be
\label{ppp8}
G_4\left(\eta,\tau\right)\sim {l\over 2\gamma \sin\varphi_1}
\left\{\cos\varphi_1 + \sin\\ln \left(\gamma\eta 
\sqrt{2\left(1 + \cos\left( 2\pi T_H \tau\right)\right)}\right)\right\} \ .
\ee
The periodicity of the Euclidean time $\tau$ is typical for thermal field
theory. From Eq.(\ref{XXVII}), in the UV limit, where $f=0$, the 
first terms in (\ref{ppp7}) and (\ref{ppp8}) vanish. On the other hand, 
in the IR limit, where $f\to \infty$, the first terms become large and the second 
terms can be neglected. Then in the IR limit, the $\tau$-dependences vanish, 
which means that the operator ${\cal O}$ becomes non-dynamical or
decouples. 
The decoupling of the operator ${\cal O}$ in the IR limit can be naturally expected 
from the RG flow.

\section{One-loop vacuum energy for scalar field in deSitter space\label{Sec2}}

In this section, we consider the renormalization group flow  
in the context of the dS/CFT correspondence\cite{strominger,hull}.
The $D$-dimensional deSitter space can be realized by embedding the dS
into 
$D+1$-dimensional flat space, whose metric is given by 
\be
\label{dS1}
ds^2 = \left(dX^1\right)^2 + \left(dX^2\right)^2 + \cdots \left(dX^{D-1}\right)^2 
+ \left(dX^D\right)^2 - \left(dX^{D+1}\right)^2 
\ee
by the constraint, 
\be
\label{dS2}
l^2 = \left(X^1\right)^2 + \left(X^2\right)^2 + \cdots \left(X^{D-1}\right)^2 
+ \left(X^D\right)^2 - \left(X^{D+1}\right)^2\ . 
\ee
The metric  (\ref{dS1}) and (\ref{dS2}) is invariant 
under the $SO(D,1)$ transformation. Then  deSitter space has an
isometry of $SO(D,1)$, which is identical with the group generated by the conformal 
transformation in the Euclidean $D-1$ space. Solving the constraint (\ref{dS2}) 
by choosing the independent coordinates as
\be
\label{dS3}
U=X^D + X^{D+1}\ ,\quad x^i={1 \over U}X^i\ , \ \left(i=1,2,\cdots, D \right)\ ,
\ee
we obtain the following metric of the dS:
\be
\label{dS4}
ds_{\rm dS}^2 = \sum_{\mu,\nu=0}^{D-1} g_{{\rm dS}\mu\nu}dx^\mu dx^\nu
= - {l^2 \over U^2}dU^2 + U^2 \left(dx^i\right)^2\ .
\ee
Here $x^0=U$. 
%%%%%%%%%%%%
If we change the coordinate $U$ by $U=\e^{t \over l}$, the metric in (\ref{dS4}) 
can be rewritten as
\be
\label{ddd1}
ds_{\rm dS}^2 = - dt^2 + \e^{2t \over l} \left(dx^i\right)^2\ ,
\ee
which has the form very similar to the metric of AdS:
\be
\label{ddd2}
ds_{\rm AdS}^2 = dz^2 + \e^{2z \over l} dx^\mu ds_\mu\ .
\ee
In case of AdS, the conformal symmetry can be realized on the surface with $z\to + \infty$. 
Then it is very natural to expect that the (Euclidean) conformal symmetry can be realized 
on the spacelike suface with $t\to +\infty$. 
The coordinate system in (\ref{ddd1}) covers the half of the
whole 
deSitter spacetime. Using the coordinate system which covers the whole
deSitter 
spacetime, the metric of the deSitter spacetime can be written as
\be
\label{ddd3}
ds_{\rm dS}^2 = - d\tau^2 + l^2\cosh^{2\tau \over l} d\Omega_{(4)}^2\ .
\ee
Here $d\Omega_{(4)}^2$ is the metric of 4d sphere with unit radius. The space-like 
surface $t\to +\infty$ in (\ref{ddd1}) corresponds to the surface $\tau\to
+\infty$ 
in (\ref{ddd3}). Then even in the metric (\ref{ddd3}), one may expect that
the conformal 
symmetry is realized on the the surface $\tau\to + \infty$. 
In the AdS/CFT set-up, the shift of $z$ in (\ref{ddd2}) corresponds to the 
scale transformation of the CFT. Then we may assume that the shift of $t$
in (\ref{ddd1}) 
should also correspond to the scale transformation. Therefore
 in a  way parallel to 
 \cite{GM}, one may consider the RG flow and c-functions  in the
framework of the 
dS/CFT correspondence. 

By using the expression of the metric (\ref{dS4}), for the two points on the dS, 
$SO(D,1)$ invariant distance can be defined by
\bea
\label{dS5}
L^2&\equiv& X_{(1)}^1 X_{(2)}^1 + \cdots + X_{(1)}^D X_{(2)}^D - X_{(1)}^{D+1} 
X_{(2)}^{D+1} \nn
&=& -{1 \over 2}U_{(1)}U_{(2)}\left(x_{(1)} - x_{(2)}\right)^2 
+ {l^2 \over 2}\left({U_{(1)} \over U_{(2)}} + {U_{(2)} \over U_{(1)}}\right)\ .
\eea
Then when  two points coincide with each other, one has $L^2=l^2$. 
In the following, we concentrate on the case of $D=5$. 
The propagator $G$ of the real scalar field $\phi$ with mass $m$ should
only 
depend on $L^2$ : $G=G\left(L^2\right)$. 
The equation determining the propagator 
\be
\label{dS6}
{1 \over \sqrt{-g_{\rm dS}}}\partial_\mu\left(g^{{\rm dS}\,\mu\nu}\partial_\nu 
G \right) - m^2 G ={1 \over \sqrt{-g_{\rm dS}}}
\delta^D\left(x_{(1)}^\mu - x_{(2)}^\mu\right)
\ee 
can be rewritten, when $x_{(1)}^\mu - x_{(2)}^\mu\neq 0$, as
\be
\label{dS7}
0=\left(-{\left(L^2\right)^2 \over l^2} + l^2\right)
{d^2 G \over d \left(L^2\right)^2} - {5 \over l^2}
{d G \over d \left(L^2\right)} - m^2 G =0\ .
\ee
With the new coordinate $z$ defined by
\be
\label{dS8}
z={1 - {L^2 \over l^2} \over 2}\ ,
\ee
Eq.(\ref{dS7}) has the following form:
\be
\label{dS9}
0=z(1-z){d^2 F \over dz^2} + \left({5 \over 2} - 5z\right)
{d F \over dz} - M^2\ .
\ee
Here $M^2\equiv m^2 l^2$. The solutions of (\ref{dS9}) are given by 
Gauss' hypergeometric functions $F(\alpha,\beta,\gamma;z)$:
\bea
\label{dS10}
G(z)&=& {\Gamma\left(\alpha_+\right) \Gamma\left(\alpha_-\right) \over 
\left(4\pi\right)^{5 \over 2}\Gamma\left({5 \over 2}\right) l^3} 
F\left(\alpha_+,\alpha_-,{5 \over 2};1-z\right)
+ {\xi \over l^3}F\left(\alpha_+,\alpha_-,{5 \over 2};z\right)\ ,\nn
\alpha_\pm &\equiv& 2 \pm \sqrt{4 - M^2}\ .
\eea
Note the above expression (\ref{dS10}) is well-known, for example, see 
\cite{EL}. 
 $z\to 0$ corresponds to $x_{(1)}^\mu - x_{(2)}^\mu\to 0$. 
The coefficient of the first term is fixed to reproduce the 
$\delta$-function in (\ref{dS6}). On the other hand the coefficient $\xi$ 
of the second term should be determined by the boundary condition. 
When $z$ is large, $G(z)$ behaves as
\bea
\label{dS11}
G(z) &\sim& {\Gamma\left(\alpha_- - \alpha_+\right) \over 
\Gamma\left(\alpha_-\right) \Gamma\left(5 - \alpha_+\right)l^3} 
\left( {\Gamma\left(\alpha_+\right) \Gamma\left(\alpha_-\right) \over 
\left(4\pi\right)^{5 \over 2}\Gamma\left({5 \over 2}\right)l^3} 
+ \xi (-1)^{-\alpha_+}\right) z^{-\alpha_+} \nn
&& + {\Gamma\left(\alpha_+ - \alpha_-\right) \over 
\Gamma\left(\alpha_+\right) \Gamma\left(5 - \alpha_-\right)} 
\left( {\Gamma\left(\alpha_+\right) \Gamma\left(\alpha_-\right) \over 
\left(4\pi\right)^{5 \over 2}\Gamma\left({5 \over 2}\right)} 
+ \xi (-1)^{-\alpha_-}\right) z^{-\alpha_-}\ ,
\eea
which corresponds to the asymptotic behavior of the scalar field $\phi$:
\bea
\label{dS12}
\phi&\sim& \alpha z^{-\alpha_+} + \beta z^{-\alpha_-}\ ,\nn
{\alpha \over \beta}&=&{{\Gamma\left(\alpha_- - \alpha_+\right) \over 
\Gamma\left(\alpha_-\right) \Gamma\left(5 - \alpha_+\right)} 
\left( {\Gamma\left(\alpha_+\right) \Gamma\left(\alpha_-\right) \over 
\left(4\pi\right)^{5 \over 2}\Gamma\left({5 \over 2}\right)} 
+ \xi (-1)^{-\alpha_+}\right) \over 
{\Gamma\left(\alpha_+ - \alpha_-\right) \over 
\Gamma\left(\alpha_+\right) \Gamma\left(5 - \alpha_-\right)} 
\left( {\Gamma\left(\alpha_+\right) \Gamma\left(\alpha_-\right) \over 
\left(4\pi\right)^{5 \over 2}\Gamma\left({5 \over 2}\right)} 
+ \xi (-1)^{-\alpha_-}\right) }\ .
\eea
Then the conformal weight $\Delta_\pm$ of the corresponding CFT primary 
fields  in dS/CFT correspondence is given by  
\be
\label{dS13}
\Delta_\pm = 4 - \alpha_\mp = \alpha_\pm = 2 \pm \sqrt{4 - M^2}\ .
\ee
One may compare Eq.(\ref{dS13}) with the expression for AdS, which is  
given by (\ref{XIVb}). Inside the square root, the sign of $l^2$ is changed. 
For AdS, if the Breitenlohner-Freedman bound $m^2\geq - {4 \over l^2}$
is not fulfilled, 
the conformal weight is real but in case of dS, in order that the 
conformal weight is real, there is an upper bound for the mass 
$m^2\leq {4 \over l^2}$.

As in Eq.(\ref{XX}), the parameters $\alpha$ and $\beta$  (\ref{dS12}) 
could be related with the coefficient $f$ (\ref{XIX}):
\be
\label{dS13b}
f={\alpha \over \pi\beta}\ .
\ee
Using (\ref{dS12}), we find the values $\xi_0$ and $\xi_\infty$ 
of $\xi$ corresponding to $f\to 0$ and $f\to \infty$, respectively:
\bea
\label{dS14}
\xi_0 &=& -(-1)^{\alpha_+} {\Gamma\left(\alpha_+\right) 
\Gamma\left(\alpha_-\right) \over 
\left(4\pi\right)^{5 \over 2}\Gamma\left({5 \over 2}\right)} \nn
&=& -{\left(3 - \alpha_+ \right)\left( 2 - \alpha_+\right) \left( 1 - \alpha_+ 
\right) \over \sin \left(\pi\alpha_+\right)\cdot 2^3\cdot 3 \pi^2} 
\e^{i\pi\alpha_+}\ ,\nn
\xi_\infty &=& - (-1)^{\alpha_-} {\Gamma\left(\alpha_+\right) 
\Gamma\left(\alpha_-\right) \over \left(4\pi\right)^{5 \over 2}
\Gamma\left({5 \over 2}\right)} \nn
&=& -{\left(3 - \alpha_+ \right)\left( 2 - \alpha_+\right) \left( 1 - \alpha_+ 
\right) \over \sin \left(\pi\alpha_+\right)\cdot 2^3\cdot 3 \pi^2} 
\e^{-i\pi\alpha_+}\ .
\eea
Then one obtains
\be
\label{dS15}
G\left(f=\infty, z=0\right) - G\left(f=0, z=0\right)
= - {\left(3 - \alpha_+ \right)\left( 2 - \alpha_+\right) 
\left( 1 - \alpha_+ \right) \over 2^2\cdot 3 \pi^2 l^3} \ .
\ee
The new parameter $\gamma$ may be defined:  
\be
\label{dS16}
\gamma\equiv \sqrt{4 - M^2}\ .
\ee
 $M^2=4$ ($m^2 = {4 \over l^2}$) corresponds to $\gamma^2=0$. 
Note the mass is not always tachyonic. 
The difference $\delta V$ of the vacuum energies corresponding to the 
two boundary conditions is given by
\bea
\label{dS17}
\delta V &=& {1 \over 2l^2} \int_0^{\gamma^2} d\tilde \gamma^2 \left\{
G\left(f=\infty, M^2=4 - \tilde\gamma^2, z=0\right) \right. \nn 
&& \left. - G\left(f=0, M^2=4 - \tilde\gamma^2, z=0\right)\right\} \nn
&=& - {1 \over 12\pi^2 l^5}
\left({\left(2 - \Delta_-\right)^3 \over 3} - {\left(2 - \Delta_-\right)^5 
\over 5}\right)\ .
\eea
The expression  (\ref{dS17}) is identical with that in the AdS bulk case 
in \cite{GM}. The only difference is the sign inside the square root in 
the expression of $\Delta_-$ (\ref{dS13}). 
$\Delta_\pm$ in (\ref{dS13}) becomes complex when $M^2>4$ or 
$m^2>{4 \over l^2}$, which may be a sign that the corresponding 
CFT is not unitary. We should also note that $\delta V$ 
(\ref{dS17}) becomes purely imaginary when $M^2>4$. 

One may now conjecture a
c-function 
as 
\bea
\label{dS18}
c_{\rm dS}&=&l^3\left(1 + \delta V\right) \nn
&=& l^3 \left( 1 - {1 \over 12\pi^2}
\left({\left(2 - \Delta_-\right)^3 \over 3} - {\left(2 - \Delta_-\right)^5 
\over 5}\right)\right)\ .
\eea
The c-function is, of course, less than $l^3$, which is the value of the 
c-function before the deformation. Then it is consistent with 
the renormalization flow or c-theorem in four dimensions is fulfilled
within dS/CFT.
The situation is really funny, we have no the explicit example of dual CFT 
but the corresponding RG flow is known (and it actually coincides with AdS
flow).
We have observed that there should be RG flow even 
for dS/CFT in the same way as for AdS/CFT case \cite{witten2}. In
\cite{GK} from CFT side it has been explicitly verified 
 that it is the same RG flow\cite{GM} interpolating between UV region,
where the scalar
operator has a dimension $\Delta_-$, and IR region, where the dimension of 
the scalar operator is $\Delta_+$.
Then the same arguments could be applied for
dS/CFT correspondence where consistent dual CFT is not constructed
yet. The geometry of dS 
will be determined by the leading order of corresponding ($1/N$?)
approximation, as in
the case of AdS, and 
the one-loop correction due to the scalar field will give next-to-leading 
order correction to 
the geometry. Thus, even in the absence of explicit example for dual CFT 
in dS$_5$/CFT$_4$ correspondence one can see that there is
indication for existance of RG flow similar
to
AdS case. 

\section {Discussion}

In summary, we found the difference of vacuum energies for boundary
conditions specified by the double-trace operator. Five-dimensional 
AdS black hole which is dual to thermal QFT and deSitter space are
considered. Using AdS/CFT (dS/CFT) correspondence such difference may be
interpreted as difference between end-points of RG flow.
As a result, the holographic c-theorem is verified beyond the leading
order approximation, generalizing the correspondent check done in
ref.\cite{GM} for pure AdS space. The remarkable fact is about dS/CFT 
where we make predictions about c-function while explicit 
(probably , non-unitary) dual CFT is not constructed yet.

In Eq.(\ref{dS10}), the propagator of the scalar field in the dS$_5$ background 
has been constructed. The corresponding propagator in the (Euclidean) AdS$_5$ 
background can be obtained by replacing the length parameter $l^2$ by 
$l^2\to -l^2$ (or ${M^2={m^2 \over l^2}\to - M^2}$). In both of the dS and 
(Euclidean) AdS cases, the propagator only depends on $SO(5,1)$ invariant 
distance $L^2$ (\ref{dS5}) or $z$ (\ref{dS8}). In case of AdS$_5$, 
one can choose the boundary, where  dual CFT lives, 
to be flat 4d surface, 4d sphere, or 4d hyperboloid. Such a choice of the 
boundary can be transformed to each other by the $SO(5,1)$ transformation. 
Since the scalar propagator is invariant under the transformation, the 
results, say for $\delta V$ in \cite{GM}, do not depend on the choice of 
the boundary. 

As a final remark let us note that it would be really interesting to
extend this discussion for the situation when other bulk fields present
(five-dimensional gauged supergravity). 

Note that recently similar questions with account of RG flow for bulk masses were 
addressed in \cite{nolland}.

\section*{Acknowledgments}

The work by S.N. is supported in part by the Ministry of Education, 
Science, Sports and Culture of Japan under the grant n. 13135208 and
that by S.D. O.-by AEI, Golm.
S.N. is grateful to the Yukawa Institute for Theoretical Physics, Kyoto 
University and S.D.O. is grateful to H. Nicolai and
Albert-Einstein-Institut, Golm for warm hospitality.
We acknoweledge helpful discussions with K. Behrndt, G. Gibbons 
and M. Cvetic as well as with participants of 
 the YITP workshop YITP-W-01-15 on 
"Braneworld-Dynamics of spacetime boundary".

\end{document}